\documentstyle[prl,aps,twocolumn]{revtex}

\input epsf

\begin{document}

\wideabs{
\title{
Producing Bose condensates using optical lattices
}
\author
{
Maxim Olshanii$^{1}$\cite{e-mail:maxim} and David Weiss$^{2}$\cite{e-mail:david}
}
\address
{
$^{1}$Department of Physics \& Astronomy,
University of Southern California, Los Angeles, California 90089-0484, USA
\\
$^{2}$Department of Physics, The Pennsylvania State University,
University Park, PA 16802, USA }

\date{\today}
\maketitle
\begin{abstract}
We relate the entropies of ensembles of atoms in optical lattices
to atoms in simple traps. We then determine which ensembles of
lattice-bound atoms will adiabatically transform into a Bose
condensate. This shows a feasible approach to Bose condensation
without evaporative cooling.
\end{abstract}
}

There is considerable interest in combining weakly interacting
Bose Einstein condensates (BECs) \cite{Stringari} with optical
lattices \cite{Deutsch}. The BEC-lattice combination is a dream
model system: interactions among atoms in the BEC can be simply
parameterized, and the optical lattice potential is exactly
calculable. Work to date, both experimental and theoretical, has
started with a BEC, and then put it in a lattice. For example,
BECs in 1D lattices have been used to make a pulsed atom laser
\cite{Kasevich} and to demonstrate four wave mixing
\cite{Phillips} and squeezing \cite{Kasevich2} of atoms. The
superfluid-Mott insulator transition has been predicted
\cite{Zoller} and observed \cite{Hansch} with BEC in a 3D
lattice. In this paper, we theoretically address the converse
problem. Atoms that have never been Bose condensed start out
deeply bound in a 3D optical lattice. The optical lattice is
adiabatically removed, so that the atoms are left in either a
flat-bottom or a harmonic trap. We determine, via {\it entropy
comparison}, those initial conditions for which the final state
is a BEC. Apart from the insight this provides for the types of
experiments performed to date, it also presents a strategy for
achieving BEC without evaporative cooling. This novel approach to
BEC could be orders of magnitude faster than existing techniques.

We consider only initial lattice distributions with either one or
zero atoms per site. These are the most likely distributions after
laser cooling in a lattice, because photoassociative collisions
efficiently remove pairs of atoms at the same site \cite{DePue}.
We assume that atoms have been optically pumped and are thermally
distributed among lattice site vibrational levels, the result of,
for instance, 3D Raman sideband cooling in an optical lattice
\cite{Vuletic} \cite{Han}.

To realize these initial lattice distributions, we assume that
atoms are initially trapped with no coupling between the sites
and an infinitely strong in-site repulsion. Each site is
represented by a three-dimensional harmonic oscillator of
frequency $\omega$. We also assume a site-dependent energy offset
$\epsilon_{\bf i} - \frac{3}{2} \hbar\omega$ of the bottom of
each harmonic oscillator
\begin{eqnarray}
\hat{H}_{l} =
\sum_{{\bf i}}\,\sum_{\{j\}}\,
\left(\epsilon_{\bf i} + \hbar\omega\!\sum_{\alpha=x,y,z}\!\! j_{\alpha}\right)\,\, \hat{n}_{{\bf i},\{j\}}
+
\tilde{\Theta}(\hat{n}_{\bf i})
\quad,
\end{eqnarray}
where the operator $\hat{n}_{{\bf i},\{j\}} \equiv
\hat{a}^{\dagger}_{{\bf i},\{j\}}\hat{a}_{{\bf i},\{j\}}$
corresponds to the number of atoms in the ${\bf i}$'th site in the
``in-site'' state given by the $\{j\} \equiv \{j_x,\,j_y,\,j_z\}$
set of the ``Cartesian'' quantum numbers, the operator
$\hat{n}_{\bf i} \equiv \sum_{\{j\}}\,\hat{n}_{{\bf i},\{j\}}$
gives the total number of atoms in the ${\bf i}$'th site, the
function
\begin{eqnarray}
\tilde{\Theta}(n) \equiv
\left\{
 \begin{array}{ll}
 0, & \mbox{for } n = 0,\,1
 \\
 +\infty & \mbox{for } n \ge 2
 \end{array}
\right.
\end{eqnarray}
provides the infinite in-site repulsion, and
$\hat{a}^{\dagger}_{{\bf i},\{j\}}$ ($\hat{a}^{\dagger}_{{\bf
i},\{j\}}$) is an operator creating (annihilating) a bosonic atom
in the ${\bf i}$'th site in the ``in-site'' state $|\{j\}\rangle$.

In what follows we will use a Grand-Canonical expression for
the entropy:
$
S = - Tr[\ln(\hat{\rho})\, \hat{\rho}]\quad,
$
where
$
\hat{\rho}
=
Z^{-1}\,\exp(-\beta(\hat{H}-\mu\hat{N}))
$
is the Grand-Canonical $N$-body density matrix of the system
normalized to unity,$Z=Tr[\exp(-\beta(\hat{H}-\mu))]$ is the partition function,
the chemical potential $\mu$ is chosen to provide the actual total number of
particles $N$:
$
Tr[\hat{N}\,\hat{\rho}] = N \, ,
$
$\hat{N}$ being the total-number-of-particles operator, and
$\beta = 1/T$ is the inverse temperature corresponding to the
total energy $E$ of the system:
$
Tr[\hat{H}\,\hat{\rho}] = E
$.

{\it Generic lattice: entropy.} The total entropy of a
lattice is the sum of the entropies of individual sites:
\begin{eqnarray}
S_{l} = \sum_{\bf i} \tilde{s}_{\bf i} \, ,
\end{eqnarray}
that according to expression for the entropy given above can be shown to be
\begin{eqnarray}
\tilde{s}_{\bf i} = \ln(\tilde{z}_{\bf i}) + \beta(e_{h.o.}(d) - \mu_{\bf i})\bar{n}_{\bf i} \, .
\end{eqnarray}
Here $\mu_{\bf i} = \mu - \epsilon_{\bf i}$ is the effective chemical potential,
$
\tilde{z}_{\bf i} = 1 + e^{\beta\mu_{\bf i}} z_{h.o.}(d)
$
is the site's partition function, and
$
z_{h.o.}(d) = (1-e^{-\beta\hbar\omega})^{-d}
$
and
$
e_{h.o.}(d) = d\, \hbar\omega \,(e^{\beta\hbar\omega} - 1)^{-1}
$
are respectively, the partition function and energy of a single
$d$-dimensional harmonic oscillator.

For a given temperature and chemical potential the total number
of particles in the lattice is
$
N = \sum_{\bf i} \bar{n}_{\bf i} \, ,
$
where \begin{eqnarray}
\bar{n}_{\bf i} = \frac{e^{\beta\mu_{\bf i}}
z_{h.o.}(d)}{1+e^{\beta\mu_{\bf i}} z_{h.o.}(d)}
\end{eqnarray}
is the mean occupation of the site.

{\it Flat lattice: entropy.} Here we will assume that all
the lattice sites have the same energy offset and the number of
available sites is $N_{s}$:
\begin{eqnarray}
\epsilon_{\bf i} = 0 \quad \forall {\bf i}
\in [-N_{s}^{\frac{1}{3}}/2;\,+N_{s}^{\frac{1}{3}}/2]^3 \, .
\end{eqnarray}
The computation of the entropy is lengthy but straightforward. It
turns out that the entropy per particle $s_{l} \equiv S_{l}/N$
(where here and below the index $l$ stands for ``lattice'') is a
universal function of the mean lattice site occupation, $ \bar{n}
= \frac{N}{N_{s}} $ (the same for each site), and the probability,
$w_0$, that the ground vibrational state of a given site is
occupied if the site itself is occupied (also the same for each
site). The entropy-per-particle is
\begin{eqnarray}
s_{l}(w_0,\bar{n}) = s_{l}^{T=0}(\bar{n}) + s_{h.o.}(w_0, 3) \, ,
\end{eqnarray}
where
\begin{eqnarray}
s_{l}^{T=0}(\bar{n}) = \bar{n}^{-1}\, \left[(1-\bar{n})\ln(1/(1-\bar{n})) + \bar{n}\ln(1/\bar{n})\right]
\end{eqnarray}
is the entropy-per-particle in the same lattice but at zero
temperature \cite{canonical}, and
\begin{eqnarray}
&&
s_{h.o.}(w_0, d) = \frac{d}{w_0^{1/d}} \times
\\
&&\quad
\left[
 w_0^{1/d}\ln(1/w_0^{1/d})
 +
 (1-w_0^{1/d})\ln(1/(1-w_0^{1/d}))
\right]
\nonumber
\end{eqnarray}
is the entropy of a single one-particle $d$-dimensional harmonic
oscillator whose ground state is occupied with a probability
$w_0$. Thermal equilibrium is assumed.

{\it Lattice with a harmonic envelope: entropy.}
Consider now a lattice whose site ground state energies
harmonically depend on the site position:
\begin{eqnarray}
\epsilon_{\bf i} = \frac{m\Omega^2 l^2}{2} i^2 \quad,
\end{eqnarray}
where $l$ is the distance between the sites assumed to form a cubic lattice.
The computation of the harmonic lattice entropy is pretty analogous
to the one for the flat lattice. It leads to
\begin{eqnarray}
s_{l}(w_0,\bar{n}) = s_{l}^{T=0}(\bar{n}) + s_{h.o.}(w_0, 3)  \, ,
\end{eqnarray}
where
$
\bar{n} = Tr[\hat{n}_{{\bf i} = {\bf 0}}\, \hat{\rho}]
$
now denotes the site occupation in the center of the lattice,
\begin{eqnarray}
&&s_{l}^{T=0}(\bar{n}) = \frac{5\mbox{Li}_{5/2}(-q(\bar{n}))}{2\mbox{Li}_{3/2}(-q(\bar{n}))}
                       - \ln(q(\bar{n}))
\\
&&q(\bar{n}) = \frac{\bar{n}}{1-\bar{n}}
\end{eqnarray}
is the entropy-per-particle at zero temperature,
and the $d$-dimensional harmonic oscillator entropy $s_{h.o.}(w_0, d)$ is given above. Here
$\mbox{Li}_{\eta}(x) \equiv \sum_{j=1}^{\infty} x^j/j^{\eta}$ is the polylogarithmic function.
In the course of the above calculation we implicitly assumed that the site occupation
was a slow function of the site index, and thus the summation over the sites was replaced
by an integral.

{\it A general power-law trap: entropy at $T_c$.} At the end of
the thought transfer process atoms are confined in a smooth trap.
In what follows, we will assume that many trap levels are
populated, so that the semiclassical approximation applies.
Within the semiclassical approximation, the number of
(non-condensed) particles and the total entropy are given by
\begin{eqnarray}
&&N^{\prime}_{t} = \int d^3{\bf r}\, d^3{\bf p}\, w({\bf r}, {\bf p})
\\
&&S_{t} = \int d^3{\bf r}\, d^3{\bf p}\, \sigma({\bf r}, {\bf p})
\end{eqnarray}
respectively, where the expressions
\begin{eqnarray}
w({\bf r}, {\bf p})
&=& (2\pi\hbar)^{-3} \sum_{n=0}^{\infty} n \, p(n|{\bf r}, {\bf p})
\nonumber
\\
\displaystyle
   &=& (2\pi\hbar)^{-3} \frac{1}{e^{\beta(h_{t}({\bf r}, {\bf p})-\mu)} -1}
\end{eqnarray}
and
\begin{eqnarray}
\sigma({\bf r}, {\bf p})
&=& (2\pi\hbar)^{-3}
\sum_{n=0}^{\infty} -\ln(p(n|{\bf r}, {\bf p})) \, p(n|{\bf r}, {\bf p})
\nonumber
\\
\displaystyle
&=& (2\pi\hbar)^{-3}
\left[
\ln(z({\bf r}, {\bf p}))
 + \frac{\beta(h_{t}({\bf r}, {\bf p})-\mu)}
 {e^{\beta(h_{t}({\bf r}, {\bf p})-\mu)} -1}
\right]
\end{eqnarray}
are the particle- and entropy phase-space densities, and
\begin{eqnarray}
&&p(n|{\bf r}, {\bf p})
= z({\bf r}, {\bf p})^{-1} e^{-\beta(h_{t}({\bf r}, {\bf p})-\mu)n}
\\
&&z({\bf r}, {\bf p})
= \sum_{n=0}^{\infty} e^{-\beta(h_{t}({\bf r}, {\bf p})-\mu)n}
\end{eqnarray}
are the distribution of the number
of particles in a volume $(2\pi\hbar)^3$ phase-space cube
centered at a point $({\bf r}, {\bf p})$,
and the corresponding partition function.
As has been shown in
\cite{Bagnato}, Bose-Einstein condensation occurs as soon as the
total number of particles exceeds the upper bound for the number
of the non-condensed particles $N^{\star} \equiv
N^{\prime}\Big|_{\mu=0}$ given by its value at zero chemical
potential \cite{bottom}:
\begin{eqnarray}
N \ge N^{\star} \Longleftrightarrow \mbox{BEC}
\label{eq:BEC_N-criterion}
\end{eqnarray}
Below we will reformulate this condition in terms of the entropy
per particle $s_{t} = S_{t}/N$.

Consider a general power-law trapping potential
\begin{eqnarray}
U({\bf r}) = \sum_{\alpha = x,y,z} a_{\alpha} (r_{\alpha})^{\lambda_{\alpha}}
\end{eqnarray}
After lengthy but straightforward calculations one can show that in this case the
critical entropy per particle assumes a universal form
\begin{eqnarray}
s_{t}^{\star} = \frac{S_{t}^{\star}}{N^{\star}} = \frac{(\eta+2) \zeta(\eta+2)}{\zeta(\eta+1)}
\end{eqnarray}
where $ \eta = 1/2 + \sum_{\alpha = x,y,z} (\lambda_{\alpha})^{-1}
$ and $\zeta(x)$ is Rieman zeta-function. For the cases of a
flat-bottom box and three-dimensional harmonic oscillator, $\eta$
is equal to $1/2$ and $2$ respectively, leading to
\begin{eqnarray}
 \begin{array}{lclclcl}
 \mbox{3D box:} &\quad& s_{t}^{\star} &=& \frac{5\, \zeta(5/2)}{2\, \zeta(3/2)} &=& 1.284\ldots
 \\
 &&&&&&
 \\
 \mbox{3D HO:} &\quad & s_{t}^{\star} &=& \frac{4\, \zeta(4)}{\zeta(3)} &=& 3.602\ldots
 \end{array}
\end{eqnarray}
Finally, assuming the trap has a positive fixed-$N$ heat
capacity, $C_{N} = \frac{\partial}{\partial T}\Big|_{N} E = T
\frac{\partial}{\partial T}\Big|_{N} S > 0$ and the critical
number of particles monotonically increases with temperature,
$\frac{\partial}{\partial T} N^{\star} > 0$, one can show that
the BEC condition (\ref{eq:BEC_N-criterion}) is equivalent to the
following condition on the entropy per particle:
\begin{eqnarray}
s_{t} \le s_{t}^{\star} \Longleftrightarrow \mbox{BEC}
\label{eq:BEC_s-criterion}
\end{eqnarray}
This form of the BEC criterion is particularly useful for
analysis of adiabatic processes.

\begin{figure}
   \leavevmode
   \centering
   \parbox{4cm}
   {  
   \begin{center}
   \epsfxsize=0.42\textwidth
                          \epsfbox{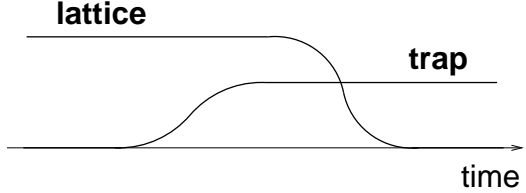}
   \end{center}
   }  
%
\caption { An artist's view of the lattice-to-trap adiabatic
transfer process. } \label{fig:process}
\end{figure}

{\it Lattice-to-trap transfer process.}
Consider the following
process. First, atoms are loaded into the optical lattice and
thermalized. Next an external trap is built around the lattice.
Finally, the original lattice is removed, and the state of the
resulting gas is detected.
The
process is schematically depicted at Fig.\ref{fig:process}.
The transfer from lattice to trap is assumed to be
thermodynamically adiabatic, and therefore the entropy of the
system is conserved throughout the process. In particular, the
entropies of the final trap and original lattice must be the
same: $S_{t} = S_{l}$. Since the number of particles $N$ is also
conserved, the above conservation law can be reformulated as
\begin{eqnarray}
s_{t} = s_{l}
\end{eqnarray}
where $s= S/N$ is the entropy per particle. Consider
a two-dimensional plane whose
coordinates are given by the peak lattice
site occupation $\bar{n}$ and ground state fraction in the
occupied sites $w_{0}$. The no-BEC vs. BEC transition line in this plane will be thus given
by an implicit equation
\begin{eqnarray}
s_{l}(\bar{n},w_{0}) = s_{t}^{\star}\quad,
\end{eqnarray}
where $s_{l}(\bar{n},w_{0})$ is the entropy per particle in the
original lattice, and $s_{t}^{\star}$ is the critical entropy per
particle for the given trap.

\begin{figure}
   \leavevmode
   \centering
   \parbox{4cm}
   {  
   \begin{center}
   \epsfxsize=0.42\textwidth\epsfbox{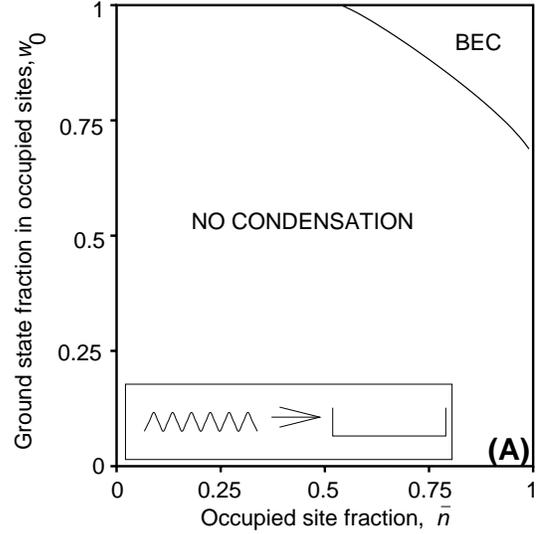}
   \end{center}
   }  
   \parbox{4cm}
   {  
   \begin{center}
   \epsfxsize=0.42\textwidth\epsfbox{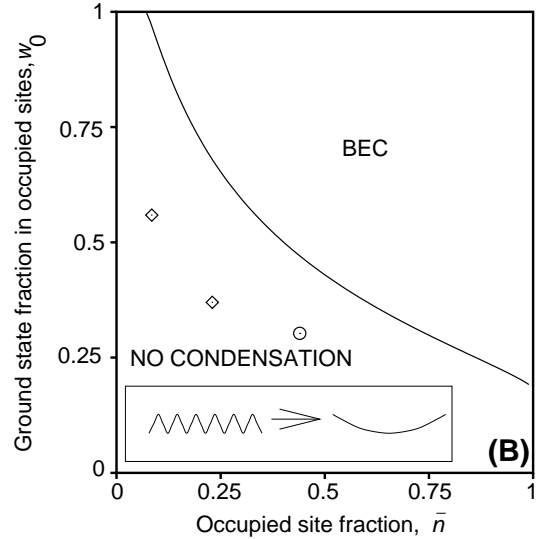}
   \end{center}
   }  
%
\caption { Phase diagram for flat-lattice-to-trap adiabatic
transfer. The final traps are (a) flat-bottom box and (b) 3D
spherically symmetric harmonic oscillator. The experimental points
correspond to results in references [8] (circle) and [10]
(diamonds). } \label{fig:box_to_trap}
\end{figure}

\begin{figure}
   \leavevmode
   \centering
   \parbox{4cm}
   {  
   \begin{center}
   \epsfxsize=0.42\textwidth\epsfbox{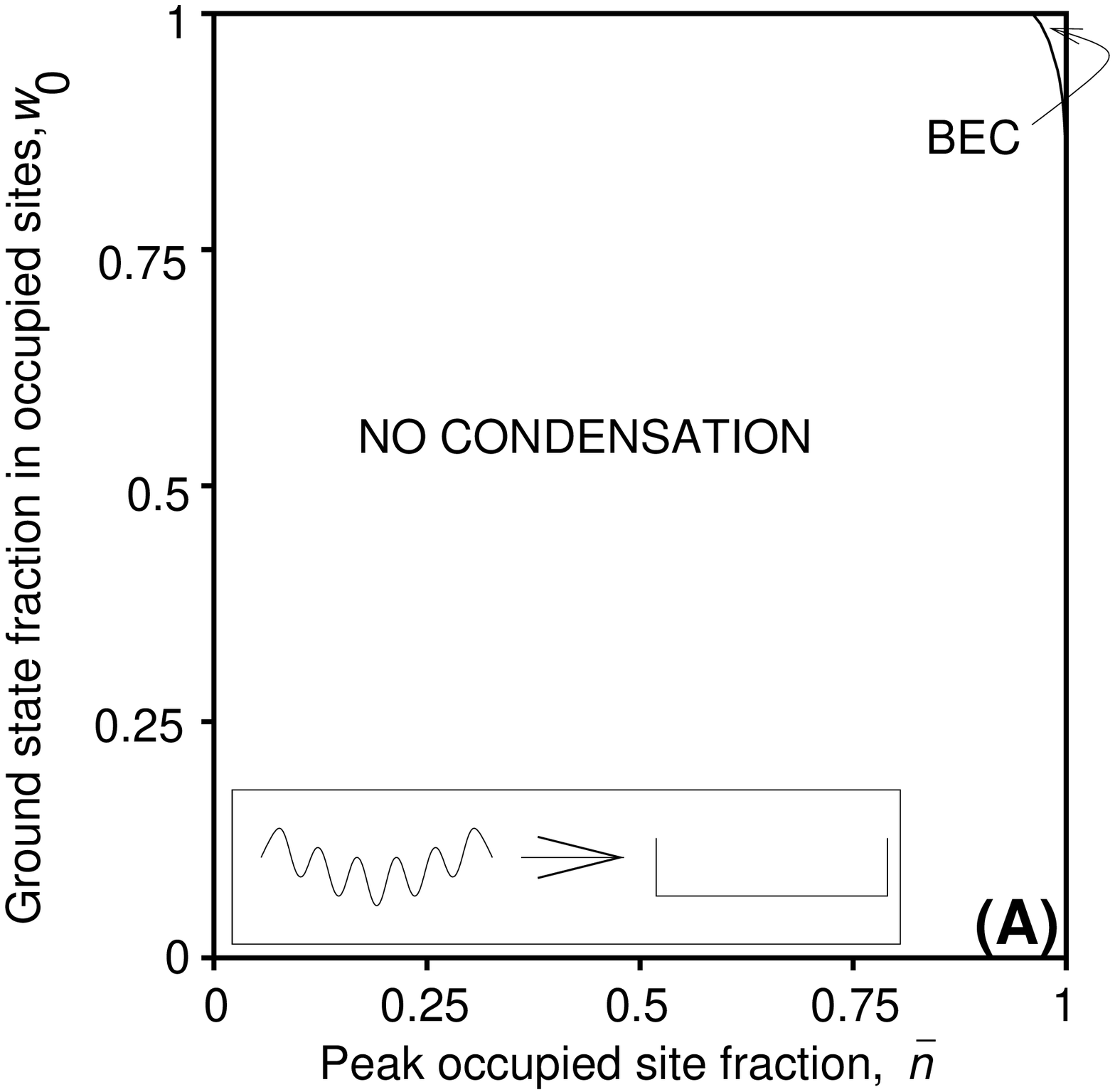}
   \end{center}
   }  
   \parbox{4cm}
   {  
   \begin{center}
   \epsfxsize=0.42\textwidth\epsfbox{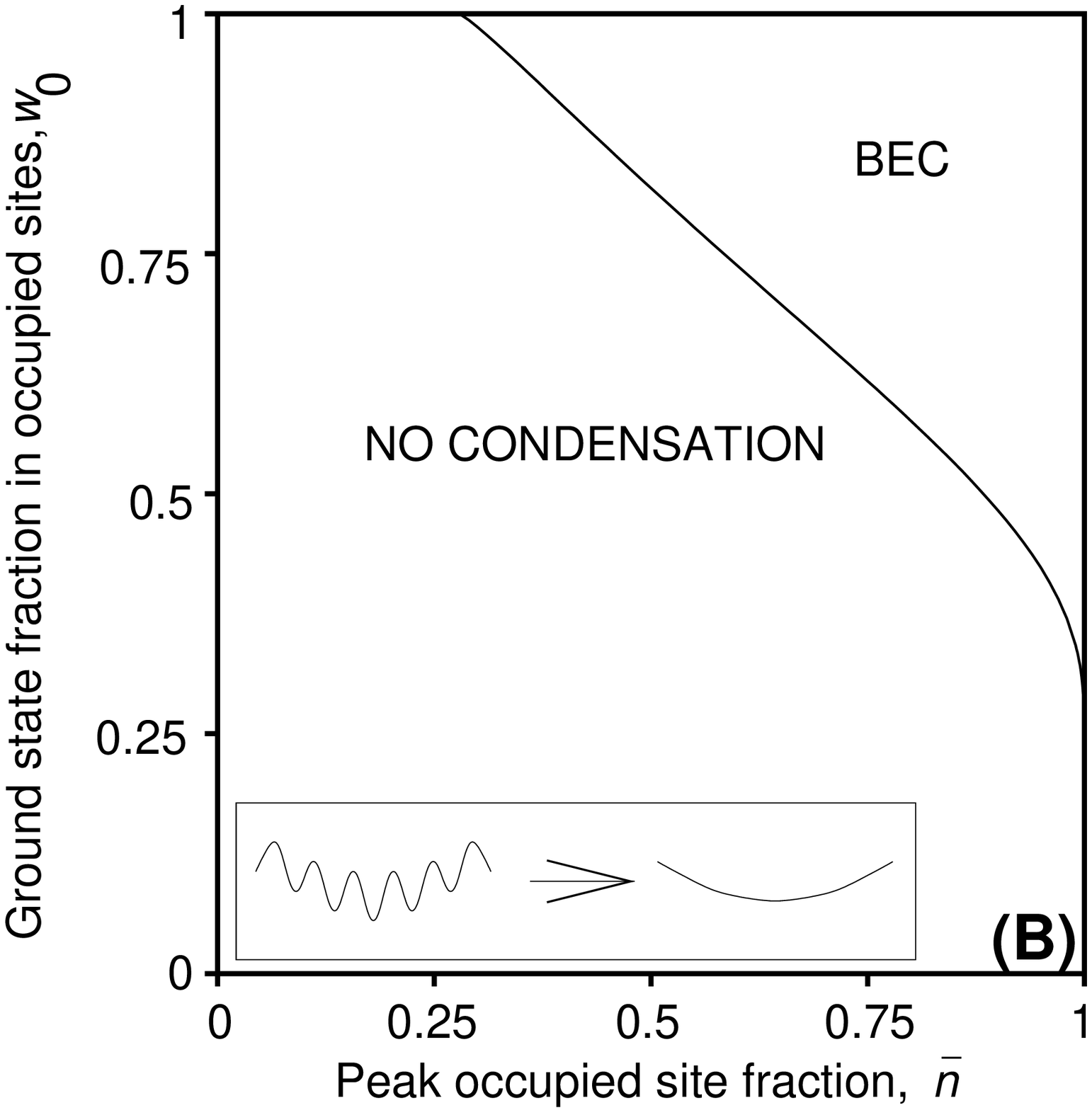}
   \end{center}
   }  
\caption
{
Phase diagram for harmonic-lattice-to-trap adiabatic transfer.
The final traps are (a) flat-bottom box and (b) 3D spherically symmetric
harmonic oscillator.
}
\label{fig:HO_lattice_to_trap}
\end{figure}

The gas in the final trap will (not) be a BEC if the lattice
entropy is below (above) the critical value $s_{t}^{\star}$. The
lattice and critical trap entropies were explicitly computed
above for two particular cases each. The four resulting phase
diagrams are shown in Figs. \ref{fig:box_to_trap} and
\ref{fig:HO_lattice_to_trap}.
In a harmonic trap, the peak phase space density exceeds the average
phase space density. Because the peak phase space density determines the
BEC threshold \cite{Ketterle}, BEC is more readily obtained when lattice
atoms are transferred to a harmonic trap than to a flat-bottom trap
(compare Figures 2B and 3B to 2A and 3A). When the initial distribution
in the lattice is flat (Figures 2A and 2B), the initial bulk phase space
density is constant across the sample. There is then more to be gained
by changing the bulk shape of the trap than when the initial phase space
density is already non-uniform  (Figures 3A and 3B). The combination of
these effects qualitatively explain why the flat lattice to harmonic
trap scenario is the most promising approach to BEC (Figure 2B).
Fortunately, the photoassociative collisions that limit the
average site occupation to one half also flatten the large scale
distribution \cite{DePue}. Existing techniques can probably
produce transient site occupations of several, which will
naturally lead to approximately flat top distributions after
laser cooling. Some published experimental results for laser
cooling in 3D optical lattices are marked on Figure 2B. Further
increases in the ground state occupation are feasible \cite{Han},
especially by scattering light in the regime where the heating
from rescattered photons is minimized \cite{Wolf}. As illustrated
in Figure 2B, only a modest improvement is required to pass the
BEC threshold.

In summary, we have related the entropy of atoms distributed in
optical lattices to the entropy of atoms in harmonic and
flat-bottom traps. By entropy comparison, we have calculated BEC
phase diagrams for lattice-bound atoms. These diagrams create a
framework for considering experiments with BECs in optical
lattices and suggest a way to obtain BEC without evaporative
cooling.

%
{\bf Acknowledgments}.
This work was supported
by NSF grants {\it PHY-0070333} and {\it PHY-0137477}, and the
Packard Foundation.

\begin{thebibliography}{99}
\bibitem{canonical}
Equation 8 can also be readily derived in the large N limit of
the canonical ensemble, where $S_{l}^{T=0}(\bar{n}) = \ln
\Gamma$, and $\Gamma$ is the number of ways N atoms can be
distributed among $N_{s}$ lattice sites. The critical site
occupation occurs when this entropy equals the entropy of a Bose
gas at the BEC threshold,$S=1.28N$. It is 0.538, which is the
uppermost point on the critical line in Figure 2A.
\bibitem{bottom} Here we assume that the trapping potential goes to zero at the
origin.
\end{thebibliography}
\end{document}